\documentclass[
aps,
prl,
reprint,
superscriptaddress,
amsmath,
amssymb,
]{revtex4-2}

\usepackage{graphicx}
\usepackage{booktabs}
\usepackage{xcolor}
\usepackage{hyperref}
\usepackage{subcaption}

\newcommand{\nrsur}{\textsc{NRSur}}
\newcommand{\xphm}{\textsc{XPHM}}
\newcommand{\xpnr}{\textsc{XPNR}}

\begin{document}

\title{On the interplay between waveform systematics and lensing signatures in gravitational-wave signals}

\author{Disha Hegde}
 \email{disha.hegde@uclouvain.be}
\affiliation{University of Louvain, Centre for Cosmology, Particle Physics and Phenomenology--CP3, Chemin du Cyclotron 2, 1348 Louvain-la-Neuve, Belgium}
 
\author{Justin Janquart}
\affiliation{University of Louvain, Centre for Cosmology, Particle Physics and Phenomenology--CP3, Chemin du Cyclotron 2, 1348 Louvain-la-Neuve, Belgium}
\affiliation{Royal Observatory of Belgium, Avenue Circulaire, 3, 1180 Uccle, Belgium}

\begin{abstract}
        The exceptional gravitational-wave event GW231123 exhibited unusually high total mass and near-extremal spins, while revealing strong waveform-dependent 
variations in its inferred source properties. Several follow-up studies favored a lensed interpretation, with reduced waveform systematics, 
raising the question of whether apparent lensing evidence could arise from waveform modeling systematics. Using GW231123-like numerical-relativity injections, we find that lensing degrees of freedom can reduce waveform discrepancies and yield non-negligible 
lensing support. However, this behavior is not systematic across injections, and cases where all waveform and lens models favor lensing while reducing waveform discrepancies are not found.

\end{abstract}

\maketitle

\section{Introduction}
\label{sec:1}
The recent observation of the exceptional gravitational-wave (GW) event
GW231123\_135430 (GW231123 hereafter)
\cite{LIGOScientific:2025rsn} is characterized by an unusually high total
mass and a preference for near-extremal spins under standard parameter
estimation (PE) analyses.
However, the inferred source properties show significant dependence on
the choice of waveform approximant, with different models yielding
noticeably different posterior distributions.
These discrepancies motivated analyses beyond the standard
interpretation. In particular, several studies investigating GW231123
under a gravitationally lensed hypothesis
\cite{LIGOScientific:2025cwb, Goyal:2025eqo,Shan:2025dcd,2025arXiv251216916C, Chakraborty:2025pxt, Hu:2025lhv, PhenomPaper,2026arXiv260721834H}
found varying support for lensing and reported reduced tensions between waveform models.
The interpretation of these results can go in two directions: 
(a) lensing is the cause of the discrepancies between waveform models
and therefore the lensing hypothesis is favored, or (b) waveform systematics 
are mitigated by the extra degrees of freedom introduced by the lensing models, leading 
to spurious support for lensing. This paper investigates, for the first time, the latter aspect. 

This question becomes increasingly relevant as the GW
catalog continues to expand. The release of the fifth Gravitational-Wave
Transient Catalog (GWTC-5) \cite{LIGOScientific:2026wfs} has increased the
number of confident compact binary coalescence observations to 390 events,
including 161 detections solely from the second part of the fourth observing run
of the LIGO--Virgo--KAGRA (LVK) Collaboration
\cite{2018LRR....21....3A}.
As detector sensitivities improve, higher signal-to-noise ratios (SNRs) and more extreme events 
will challenge waveform model accuracy. At the same time, the detection 
of novel effects such as lensing becomes increasingly 
likely~\cite{Ng:2017yiu, Li:2018prc, Xu:2021bfn, Wierda:2021upe}.
Therefore, disentangling genuine signatures of missing physics 
from waveform modeling uncertainties is essential for
reliable interpretation of future observations.

In this work, we investigate whether waveform modeling systematics can
produce false lensing signatures in GW231123-like BBH signals. We perform
a zero-noise injection--recovery study using numerical relativity (NR)
waveforms from the Simulating eXtreme Spacetimes (SXS) catalog~\cite{Scheel:2025jct}, selected to represent
the source properties inferred for GW231123 under standard unlensed
analyses and recover these signals under both unlensed and lensed hypotheses. 

The remainder of this paper is organized as follows. We first describe the injection--recovery framework, 
including the construction of the NR injections and the Bayesian inference setup. 
We then present the recovery results and conclude with a summary of our findings
and possible avenues to differentiate between genuine lensing signatures and waveform systematics in future analyses.
\section{Methods}
\label{sec:2}

We select quasi-circular, precessing BBH simulations from the SXS catalog
following the procedure described in Sec.~I of the Supplemental Material~\cite{SM}. 
The initial selection is restricted to high-spin systems with mass ratios and effective
spins consistent with the source properties inferred for GW231123. This
provides a set of candidate simulations from which we select events 
with high chances of exhibiting waveform systematics in unlensed recoveries.

Each selected SXS waveform is injected into a two-detector network consisting
of Advanced LIGO Hanford (H1) and Livingston (L1)
\cite{LIGOScientific:2014pky}. Injections are performed in zero-noise frames
and weighted using the power-spectral densities (PSDs) from the unlensed
PE analyses of GW231123~\cite{LIGOScientific:2025rsn}. 
For each injection, the luminosity distance is rescaled to achieve a target network optimal SNR of 20.

To identify simulations where waveform-model differences are expected to be
most pronounced, we quantify the disagreement between the SXS waveforms and
recovery approximants using the mismatch~\cite{PhysRevD.46.5236,PhysRevD.53.6749},
 
\begin{equation}
\label{eq:mismatch}
\mathcal{M}=1-
\max_{t_c,\phi_c,\psi,\alpha_0}
\frac{\langle h_s|h_t\rangle}
{\sqrt{\langle h_s|h_s\rangle
\langle h_t|h_t\rangle}},
\end{equation}
 
where $h_s$ and $h_t$ denote the SXS signal and template waveforms, respectively,
and $\langle \cdot | \cdot \rangle$ denotes the usual noise-weighted 
inner product~\cite{Veitch:2014wba}. The maximization is performed over the
coalescence time $t_c$ and phase $\phi_c$, the template polarization angle $\psi$, and
the initial in-plane spin orientation $\alpha_0$ of the template, following the
generalized-match procedure in Refs~\cite{Harry_2016} and~\cite{Hamilton_2021}. For the computations, we use
the L1 detector PSDs from the unlensed PE analysis of GW231123~\cite{LIGOScientific:2025rsn}.
Further details are given in Sec.~II of the Supplemental Material~\cite{SM}.
A mismatch is computed for each of the three waveforms used in this work:
\textsc{NRSur7dq4} (\nrsur)~\cite{Varma:2019csw}, \textsc{IMRPhenomXPHM-SpinTaylor} 
(\xphm)~\citep{Pratten:2020ceb,Colleoni:2024knd}, 
and \textsc{IMRPhenomXPNR} (\xpnr)~\cite{Hamilton:2025xru}.
The SXS--\nrsur \, mismatch is used to rank them, and the
40 simulations with the largest mismatches are selected, as these are expected
to show the strongest waveform systematics in unlensed recoveries.

The first 20 SXS injections with the largest mismatches (spanning $\mathcal{M}_{\rm L1}^{\rm NRSur} 
\sim 2.4\times10^{-2}$ to $0.7\times10^{-2}$) are then recovered under
both unlensed and lensed hypotheses, using three previously-introduced waveform 
approximants: \nrsur, \xphm, and \xpnr.
The unlensed analyses are performed using \textsc{bilby} \cite{Ashton:2018jfp} and
\textsc{bilby\_pipe} \cite{Romero-Shaw:2020owr}, adopting
prior distributions as described in Sec.~III of the Supplemental Material~\cite{SM}. 
The lensing analyses are done with \textsc{Gravelamps}~\cite{wright2022gravelamps} 
and we use two different lens models: an isolated point-mass (PM) lens model in 
the wave-optics regime~\cite{1998PhRvL..80.1138N,2003ApJ...595.1039T}, 
and a phenomenological lensing model in the geometric optics regime~\cite{Liu:2023ikc}. 
Priors for the PM analyses are similar to those used in Ref.~\cite{LIGOScientific:2025cwb} 
for the redshifted lens mass ($M_{l}^{z}$) and the source position ($y$), while the priors 
for the phenomenological lensing analyses are similar to those 
used in Ref.~\cite{PhenomPaper}, covering relative magnifications ($\mu_{ij}$), 
time delays $(\Delta t_{ij})$ and Morse phase indices ($n_i$). 
They are also detailed in Sec.~III of the Supplemental Material~\cite{SM}.

The remaining 20 injections are recovered under the unlensed and phenomenological-lensing hypotheses only, using  the same analysis setups, as an additional check 
on the generality of the above results. As discussed in the Results section, 
this yields no new qualitative information. We therefore neither carry out PM-lensing recoveries for 
this subset nor report the corresponding individual results in this work.
\section{Results}
\label{sec:3}

As a baseline, we recover the 40 unlensed SXS injections under the standard 
unlensed hypothesis using the three waveform approximants. Despite the large 
waveform mismatches used to select the injection set, we observe a range of 
behaviour in the unlensed recoveries, from substantial disagreements between 
approximants to negligible differences, consistent with the findings of 
Ref.~\cite{LIGOScientific:2025rsn}. We focus the detailed lensed analysis below on
the first 20 injections with the largest \nrsur \, mismatches, as described in the previous section.
All of these injections exhibit appreciable waveform-dependent discrepancies, 
defining the subset relevant for assessing whether waveform systematics can mimic 
lensing. Among these cases, \nrsur\, generally provides the most faithful recovery 
of the injected parameters, followed by \xpnr, whereas \xphm\, most frequently 
exhibits larger deviations, in line with similar studies done previously~\cite{Akcay:2025rve}.
This comparatively poorer performance of \xphm, is also not unexpected given that, 
among the three approximants considered here, it is the only one not calibrated to precession in the strong-field regime. 
Since all selected NR simulations have high spins with a spin-precession parameter (see~\cite{Gerosa_2021}) $\chi_p \geq 0.5$, 
they probe precisely a region of parameter space where this limitation may become relevant.

\begin{figure*}
    \captionsetup{justification=raggedright,singlelinecheck=false}
	\includegraphics[width=\textwidth]{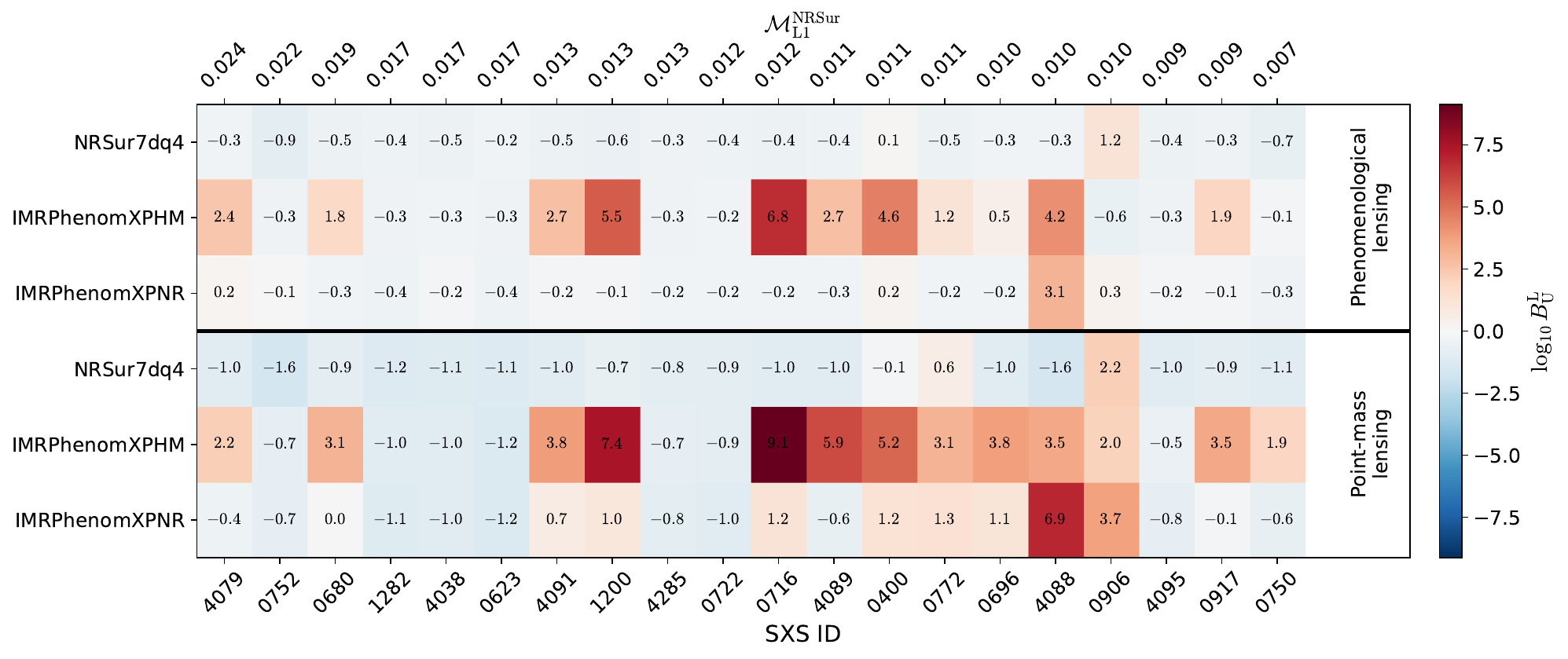}
	\caption{Logarithmic Bayes factors comparing lensed and unlensed hypotheses for the 20 SXS injections with the largest \nrsur \, mismatches, 
    which are recovered under all three hypotheses. The color scale shows 
    $\log_{10} B^{\rm L}_{\rm U}$, where positive (negative) values indicate support for the lensed (unlensed) hypothesis. The upper and lower panels 
    correspond to phenomenological and PM lensing models, respectively, with rows showing the recovery using the three waveform 
    approximants. Each column corresponds to an individual 
    SXS injection (labelled by the simulation ID suffix) ordered by decreasing mismatch $\mathcal{M}_{\rm L1}^{\rm NRSur}$ relative to \nrsur, whose 
	values are indicated in the upper axis. No clear trend between mismatch and 
    support for lensing is identified.
	}
	\label{fig:bayes_factors}
\end{figure*}

We next analyze these injections under the lensing hypothesis. We quantify the preference for lensing
using the Bayes factor $\log_{10}\mathcal{B}^{\rm L}_{\rm U, \, X}$~\cite{LIGOScientific:2025cwb}, where ${\rm X} = \{\rm phen, \, PM\}$ indicates
the phenomenological or PM lensing model, and
positive Bayes factors indicate preference for the lensed model
over the unlensed one. The Bayes factors for all injections and waveform approximants are shown in Fig.~\ref{fig:bayes_factors}.
Of the 20 injections, we find that $11$ ($13$) yield a positive Bayes factor in at least one
approximant, $4$ ($9$) in at least two, and $1$ ($2$) in all three for the phenomenological (PM) model.  A clear waveform
dependence is observed: \xphm \, produces the largest number of false-positive lensing detections, followed by \xpnr \, and then \nrsur.
This trend is consistent with the overall hierarchy of waveform inaccuracies relative to the injected SXS signals, with \xphm \, 
exhibiting the largest mismatches and PE deviations, followed by \xpnr \, and \nrsur.
In all cases where lensing is preferred, the posterior favors a two-image configuration for the phenomenological model. 
We also see that the PM model produces a larger number of false-positive lensing detections than the phenomenological model, 
likely due to the smaller Occam penalty associated to this model. Importantly, none of our injections leads to a favoring of lensing across 
lensing models and waveform approximants, as could be expected for a genuine lensed signal.

\begin{figure*}
\captionsetup{justification=raggedright,singlelinecheck=false}
\begin{subfigure}{0.48\textwidth}
\includegraphics[width=\linewidth]{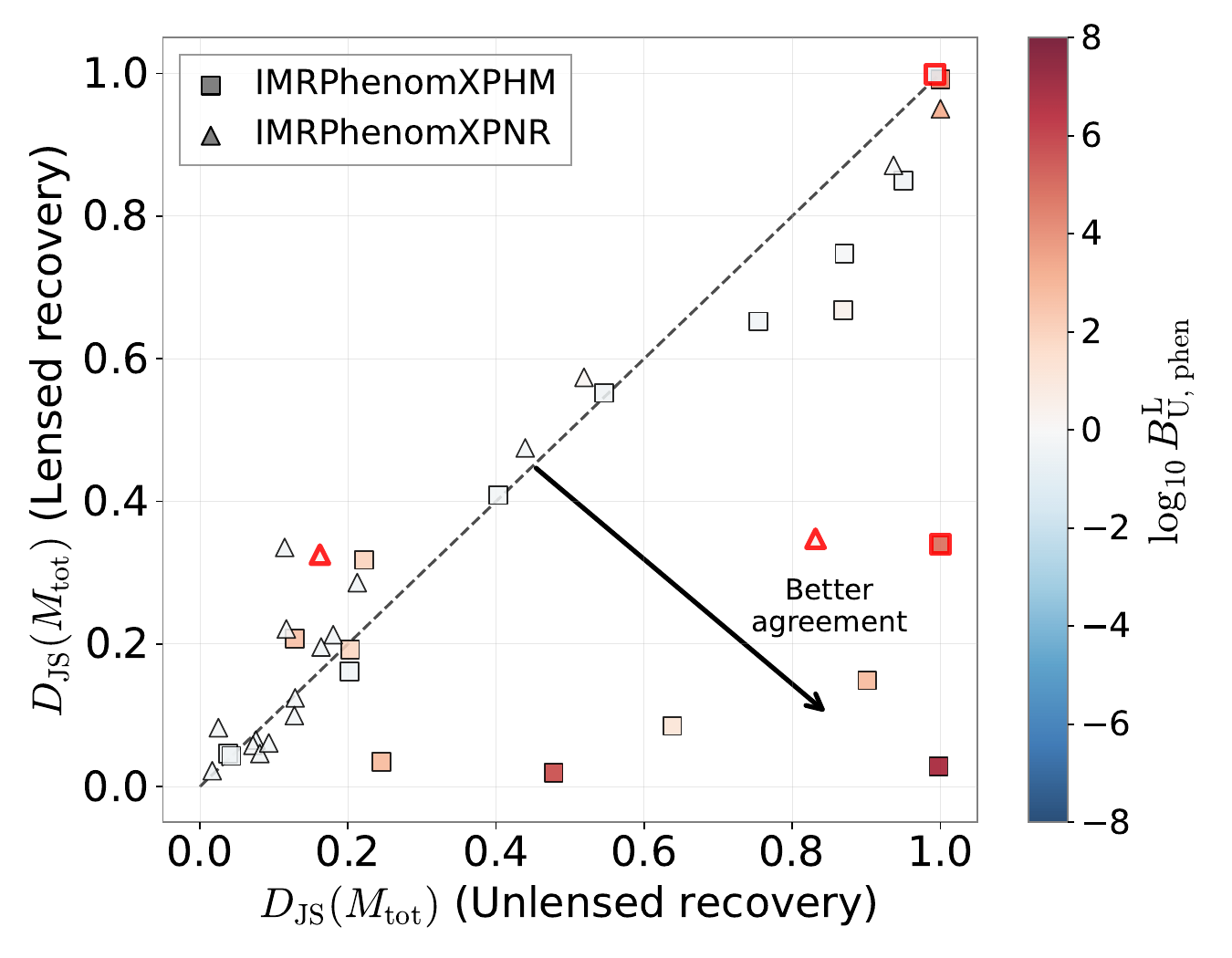}
\end{subfigure}
\hfill
\begin{subfigure}{0.48\textwidth}
\includegraphics[width=\linewidth]{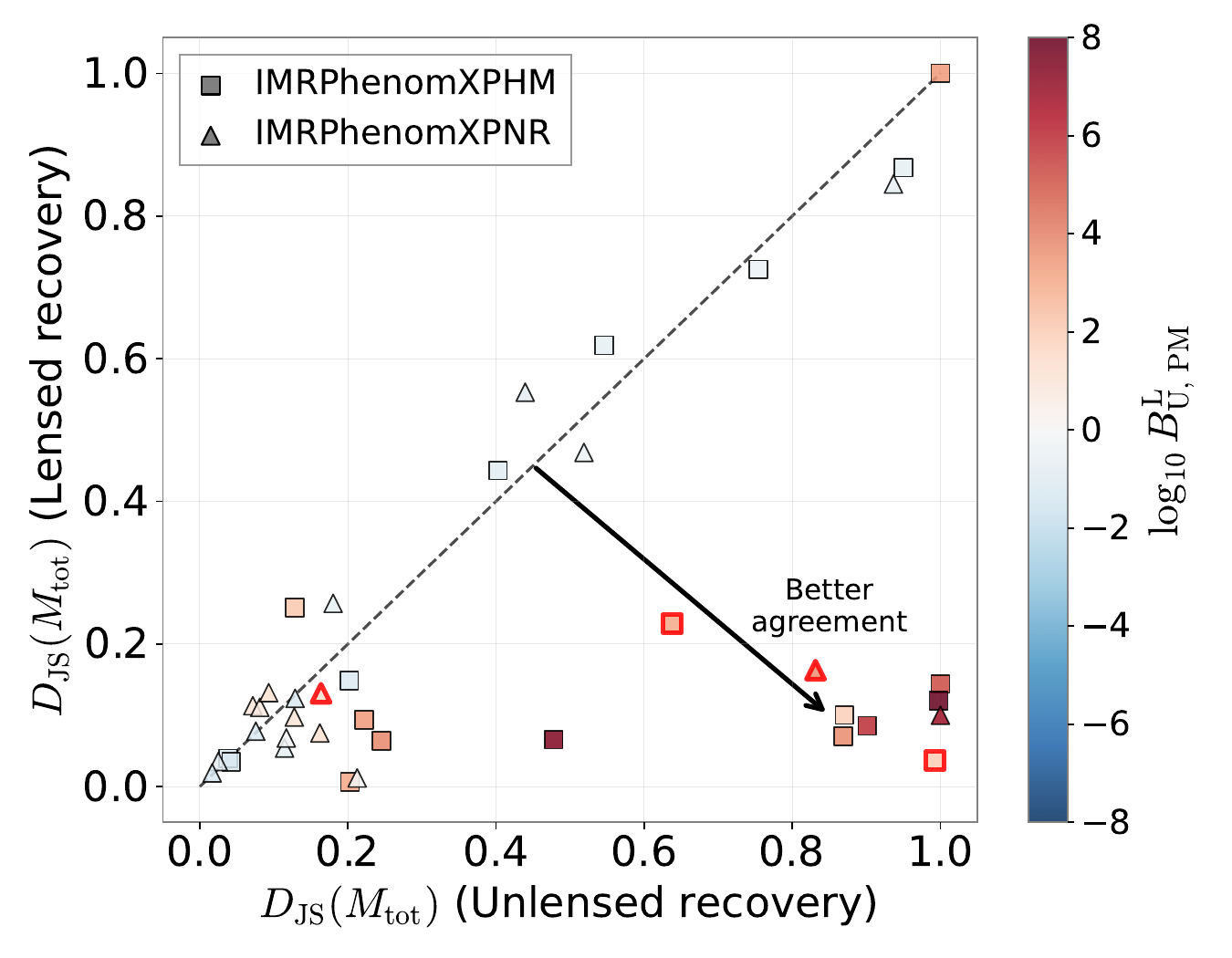}
\end{subfigure}
\caption{
Jensen--Shannon (JS) divergence between detector-frame total-mass posteriors from \nrsur \, and \xphm\,/\xpnr \, under unlensed 
(horizontal axis) and lensed (vertical axis) hypotheses. Left and right panels show the phenomenological and point-mass lens models, respectively. 
Each point represents an SXS injection, coloured by the \textsc{Phenom} lensing preference $\log_{10}\mathcal{B}^{\rm L}_{\rm U}$; red outlines mark cases where \nrsur \, 
also favours lensing. Lensing generally reduces inter-approximant differences, though not always, as indicated by points near or above the diagonal.
}
\label{fig:js_plot}
\end{figure*}

We also quantify how introducing lensing changes the agreement between waveform approximants by computing the Jensen--Shannon (JS) divergence between posterior 
distributions obtained with \nrsur\ and those obtained with \xphm\ or \xpnr, under both the unlensed and lensed hypotheses for the two lensing models, as shown in Fig.~\ref{fig:js_plot}.
Larger JS  values correspond to greater disagreement between recovered posteriors. The horizontal and vertical axes show the JS divergence under the unlensed and lensed hypotheses, 
respectively. For cases with positive lensing Bayes factors, most points lie below the diagonal, demonstrating that the additional lensing degrees of freedom reduce 
disagreements between waveform approximants by partially absorbing waveform-model differences. However, this behaviour is not universal: a small number of injections show a slight preference for lensing while remaining close to, or even above, the diagonal, 
indicating little or no improvement in the agreement between waveform approximants. Individual posterior distributions are shown in Secs.~IV--V of
the Supplemental Material~\cite{SM}.

Beyond the Bayes factors themselves, we ask whether the recovered lensing parameters in the false-positive cases describe a
physically plausible lensing scenario, or point instead to a negligible effect. Across the false-positive PM recoveries, some redshifted lens masses cluster at $M_L^z\sim1500$--$2000\,M_\odot$ with source positions $y\sim0.6$--$1.3$ (see Sec.~V of \cite{SM}),
values that would produce a non-negligible magnification ($\sim1.2-1.9$) and time-delay imprint if genuine, rather than trivially collapsing to the
unlensed limit. However, there are cases where the lensing effect would be weak despite a high Bayes factor. For example, SXS:BBH:4088 recovers $y\sim3$ for one of the three approximants,
corresponding to a magnification indistinguishable from unity, with a lensed--unlensed waveform mismatch of $6.5 \times 10^{-3}$.
 
We now examine a subset of representative injections in greater detail to investigate how false lensing evidence is coupled to waveform systematics. 
Bayes factors for all the cases are summarized in Fig.~\ref{fig:bayes_factors}; posteriors for the detector-frame total mass, mass ratio, and spin 
magnitudes $a_1$, $a_2$ under all the three hypotheses with the corresponding lens parameters are shown in the Appendix.
SXS:BBH:0400, SXS:BBH:0772, and SXS:BBH:0906 are the clearest illustration in our sample where lensing degrees of freedom absorb waveform systematics. 
PM recoveries converge on similar lens masses ($M_L^z\sim1500$--$2000\,M_\odot$) and source positions ($y\sim0.6$--$0.8$), and including lensing, particularly the PM model, 
substantially reduces inter-approximant differences in the total-mass posterior, removing bimodality where present (0772 and 0906).
On the contrary, for SXS:BBH:0400, \nrsur \, favors an effectively unlensed solution under the phenomenological model, consistent with its near-zero Bayes
factor, and the PM lens parameters, while well constrained, are recovered alongside a negative Bayes factor and hence not interpreted as evidence for lensing. 
While a systematic reduction in spins was observed for GW231123 under the lensed hypothesis
and degeneracies between lensing and spin effects exist~\cite{Liu:2023emk}, our
spin-magnitude posteriors show no common trend after including lensing. 
This likely reflects a more complex interplay between the lensing--spin-precession degeneracy and waveform-model systematics, although confirming this would require a larger injection set.
As a final example, SXS:BBH:4088 shows that reconciliation across approximants is not universal. Under the phenomenological model, \xphm \, and \xpnr \, 
favour a two-image solution, while \nrsur \, prefers the unlensed case. In the PM model, \nrsur \, and \xpnr \, recover similar lens masses ($\sim1400\,M_\odot$) 
but large source positions ($y\sim3$), implying negligible magnification; the strong \xpnr \, Bayes factor therefore likely reflects absorption of waveform systematics 
rather than a physical lensing signature. In contrast, \xphm \, favours lens parameters ($M_L^z\sim1500\,M_\odot$, $y\sim1$) consistent with a detectable lensing effect. While 
lensing partially reduces mass-posterior differences, particularly between \nrsur \, and \xpnr \,, substantial disagreement remains, and neither lens model fully reconciles 
the intrinsic parameters across all approximants.
 
Taken together, these cases demonstrate that introducing lensing degrees of freedom can reduce discrepancies between waveform approximants in intrinsic parameter 
estimation while simultaneously producing support for the lensed hypothesis. 
However, some injections exhibit substantial improvement
in the agreement between approximants, whereas others retain significant differences despite the inclusion of lensing. 
Moreover, effects are generally not consistent across waveform and lens models. 
As a check on the robustness of these findings, we repeat the unlensed and phenomenological-lensing recoveries for the next 20 SXS injections with the largest \nrsur--SXS  mismatches.
Only 6 of these 20 injections exhibit 
appreciable waveform-dependent discrepancies in the unlensed recovery in the first place. When analyzed under the phenomenological lensing hypothesis, 
none of them yield a positive Bayes factor for \nrsur \, or \xpnr, while \xphm \, does so in 4 cases, the largest being 
$\log_{10}\mathcal{B}^{\rm L}_{\rm U, \, phen}\sim2.4$. Since this second set does not add any new qualitative behaviour to the picture established above, we do 
not carry out PM-lensing recoveries for it and do not report the individual results.

\section{Conclusion}
\label{sec:4}
In this work, we have investigated whether waveform systematics in standard GW PE analyses can give rise to spurious evidence for gravitational 
lensing in GW231123-like BBH signals. 
Using zero-noise injections from the SXS catalog chosen to maximize waveform mismatches
(and chances of systematics between waveforms) and Bayesian PE under both unlensed and lensed hypotheses, we compared three waveform approximants and 
two lensing models: a phenomenological lens model and an isolated point-mass lens model.

Across studied injections, we find that allowing for lensing frequently reduces discrepancies between waveform approximants while leading to non-negligible 
support for lensing. This indicates the presence of degeneracies between 
waveform systematics and lensing signatures, indicating that careful treatment 
of waveform uncertainties will be needed in the future.
This behaviour is strongly dependent on both the waveform approximant and the specific 
injection considered, and is not consistently observed across all injections. In particular, we find that no single injection shows a uniform preference for lensing across all
waveform approximants under both lensing models, highlighting the absence of a robust, model-independent lensing signature in these cases.
Based on known degeneracies between lensing and other effects such as beyond 
general-relativity models~\cite{Mishra:2023vzo, Wright:2024mco, Gupta:2024gun} or eccentricity~\cite{Mishra:2025dpa}, similar effects are 
likely to arise for such studies. Explicitly quantifying impact for 
other scenarios is left for future work. 
However, to obtain noticeable support, we need comparatively strong 
waveform systematics, as evidenced by the absence of common lensing support 
in a waveform other than \xphm \, for the next 20 highest-mismatch injections, which show appreciable unlensed systematics in just 6 of the 20 cases.

Our study also shows possible inquiries one can conduct to differentiate between genuine lensing signatures and waveform systematics. Robust lensing support should persist 
across waveform and lens models, with consistent lens and source parameters recovered across models.
However, none of the injections we studied simultaneously shows 
(i) substantial lensing support across all waveform approximants and lens models, (ii) physically consistent lens parameters, (iii) reconciliation of intrinsic source parameters, 
and (iv) a preference for lower or uninformative spin values expected from the degeneracy 
between lensing and spin effects.
Therefore, while our work needs to be complemented by further injection studies left for future (e.g. other regions of the parameter space, 
including noise effects, looking at lensed injections recovered with imperfect waveform models), our results already suggest that
it should be possible to distinguish between genuine lensing and waveform systematics in future analyses, though care will be needed to avoid false positives.

\section{Data Availability}

Data products associated with this work, including SXS injection parameters, configuration file skeletons and analysis results are available in \cite{zenodo}.
\section{Acknowledgements}
We thank Soumen Roy for guidance on accessing and using the SXS numerical relativity simulations. 
We are grateful to Mick Wright and Jef Heynen for their invaluable help with \textsc{Gravelamps} and setting 
up the parameter-estimation runs. We also thank Mick Wright and Ania Liu for helpful discussions on the 
phenomenological lensing model.
This work was supported by the Fonds de la Recherche Scientifique (F.R.S.-FNRS) under the funding IISN-4.4501.19.
JJ also acknowledges support from the Royal Observatory of Belgium.
Computational resources have been provided by the supercomputing facilities of the Université Catholique de 
Louvain (CISM/UCL) and the Consortium des Équipements de Calcul Intensif en Fédération Wallonie Bruxelles (CÉCI) 
funded by the Fond de la Recherche Scientifique de Belgique (F.R.S.-FNRS) under convention 2.5020.11 and by the Walloon Region.
The authors are also grateful for computational resources provided by the LIGO Laboratory and supported by National Science Foundation Grants PHY-0757058 and PHY-0823459.
\appendix
\section{Posterior distributions for case studies}
\label{app:1}
\begin{figure*}
\captionsetup{justification=raggedright,singlelinecheck=false}
\includegraphics[width=\textwidth]{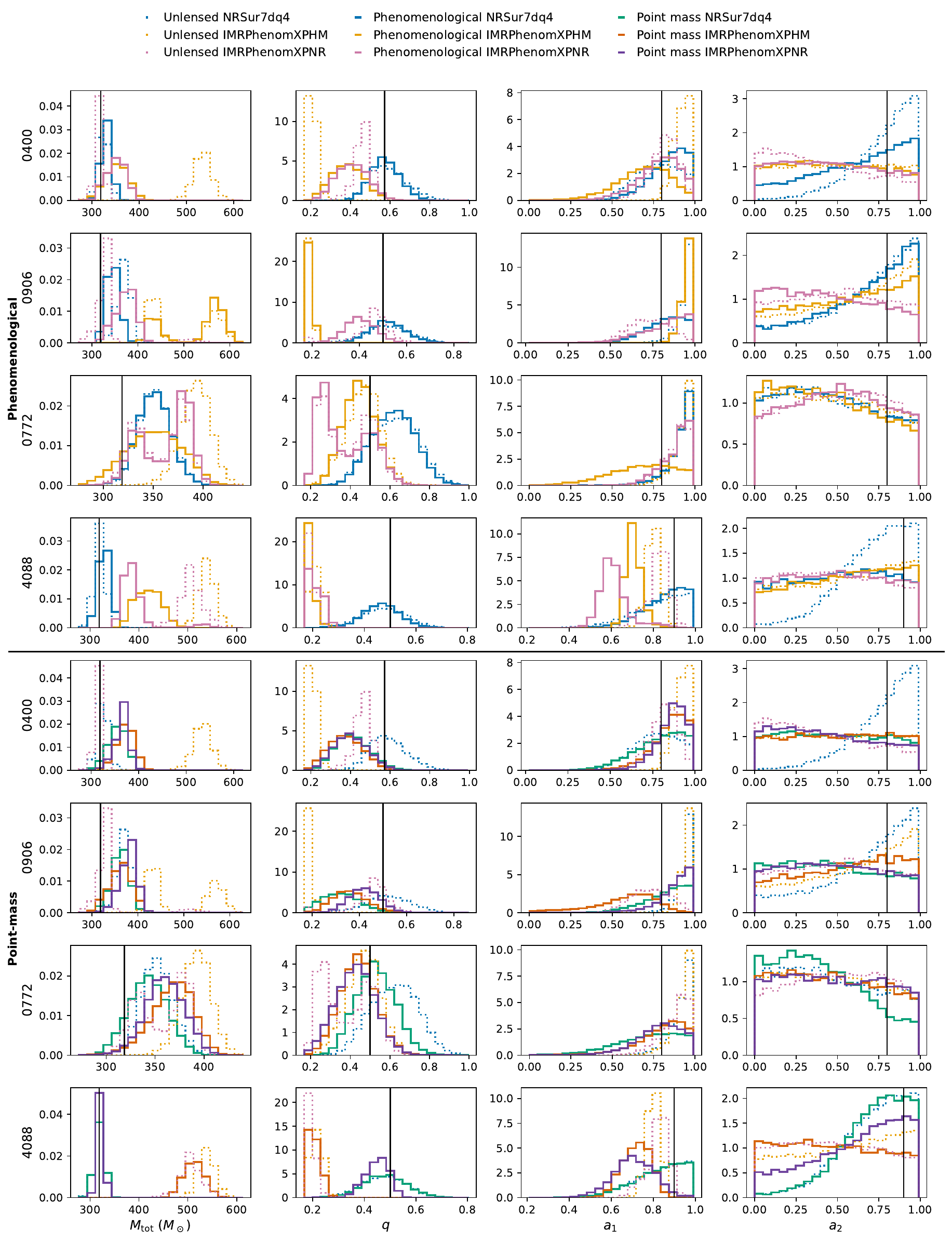}
\caption{
Posterior distributions (1D marginal histograms) for BBH parameters for the four case studies discussed in the main text. The first and last four rows show recoveries with 
the phenomenological and point-mass lensing models, respectively.}
\label{fig:posterior_case_studies}
\end{figure*}

\begin{figure*}
\captionsetup{justification=raggedright,singlelinecheck=false}
\includegraphics[width=\textwidth]{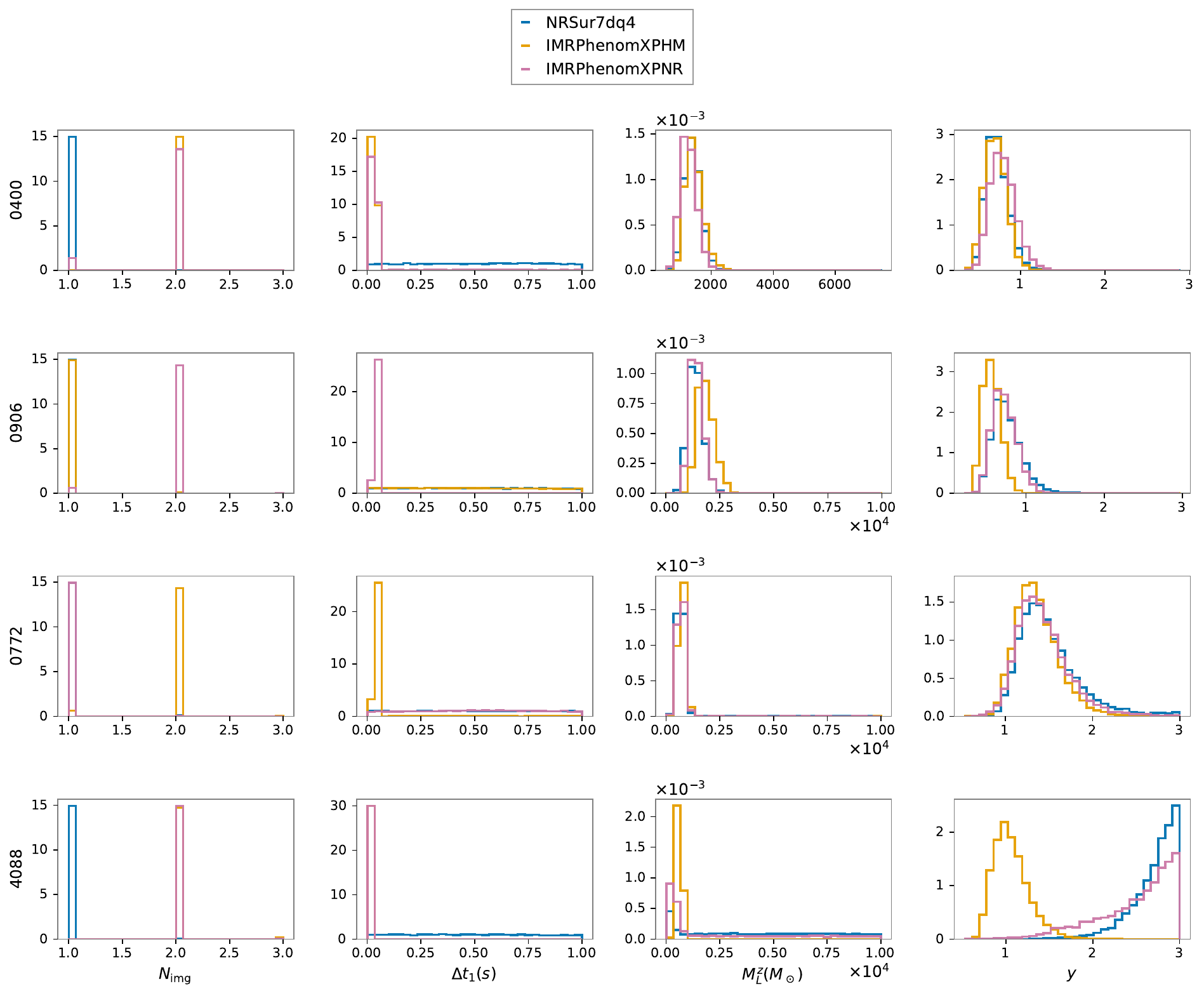}
\caption{
Posterior distributions (1D marginal histograms) of the lensing parameters for the four case studies discussed in the main text. 
Columns show, from left to right, the number of lensed images $N_{\mathrm{img}}$ and the time delay between the first two images $\Delta t_1$ inferred from the phenomenological 
lensing model, and the redshifted lens mass $M_L^{z}$ and source position $y$ inferred from the point-mass lensing model. 
For each quantity, results are shown for the three waveform approximants:
$\text{NRSur7dq4}$ (blue), $\text{IMRPhenomXPHM}$ (yellow), and $\text{IMRPhenomXPNR}$ (pink), allowing comparison of the recovered lensing configurations across waveform models.}
\label{fig:lens_posteriors}
\end{figure*}

This Appendix presents the full posterior distributions for the four representative SXS injections discussed in the main text. Fig.~\ref{fig:posterior_case_studies} shows 
the recovered intrinsic BBH parameter posteriors under the unlensed, phenomenological, and PM lensing hypotheses, illustrating the impact of introducing lensing degrees
of freedom on the agreement between waveform approximants. Fig.~\ref{fig:lens_posteriors} shows the corresponding lensing parameter posteriors for the same injections, 
providing the inferred lens configurations associated with the apparent lensing preferences discussed in the main text.

\bibliographystyle{apsrev4-2}
\bibliography{references}

\end{document}


\title{Supplemental Material:\\
On the interplay between waveform systematics and lensing signatures in gravitational-wave signals}

\author{Disha Hegde}
 \email{disha.hegde@uclouvain.be}
\affiliation{University of Louvain, Centre for Cosmology, Particle Physics and Phenomenology--CP3, Chemin du Cyclotron 2, 1348 Louvain-la-Neuve, Belgium}
 
\author{Justin Janquart}
\affiliation{University of Louvain, Centre for Cosmology, Particle Physics and Phenomenology--CP3, Chemin du Cyclotron 2, 1348 Louvain-la-Neuve, Belgium}
\affiliation{Royal Observatory of Belgium, Avenue Circulaire, 3, 1180 Uccle, Belgium}

\maketitle

\section{Numerical relativity injections}
\label{sec:sm_nr}

This section provides additional details on the selection of the
numerical-relativity (NR) injections used throughout this work.
The simulations are selected from  quasi-circular,
precessing binary black-hole (BBH) waveforms in the SXS catalog
\cite{Scheel:2025jct} to represent the source properties inferred for
GW231123. They are first restricted to high-spin configurations by
requiring $\chi_1 > 0.7$ and $\chi_2 > 0.6$.
This is in line with the high-spin nature of GW231123, 
and also ensures a regime favorable to waveform systematics~\cite{Akcay:2025rve}.
Within this subset, we select simulations with $q\in[0.5,1]$, where
$q=m_2/m_1\le1$. We further require
$
|\chi_{\rm eff}-\chi_{\rm eff}^{\rm target}|<0.4 ,
$
where $\chi_{\rm eff}^{\rm target}=0.31$, chosen to approximately match the
GW231123 source properties. We adopt this relatively high tolerance on $\chi_{\rm eff}$ to retain a sufficient number of simulations after imposing the
other selection criteria, while encompassing the range of $\chi_{\rm eff}$ values recovered for GW231123 with different waveform approximants.
These conditions result in 282 candidate simulations.

Apart from the mass ratio, spin configuration and luminosity distance,
all injections share a common set of source parameters derived from
the median posterior values obtained from the unlensed
\textsc{NRSur7dq4}
analysis of GW231123.
These parameters are summarized in Table~\ref{tab:common_pars}.
The luminosity distance of each injection is then rescaled to produce a
network optimal signal-to-noise ratio (SNR) of 20.

\begin{table}[h]
\caption{Parameters common to all injections.}
\label{tab:common_pars}
\begin{ruledtabular}
\begin{tabular}{ll}
Parameter & Value\\
\hline
$M_{\rm tot} \,(M_\odot)$ & $318.762$\\
$t_c~({\rm s}) $ & $1384782888.618$\\
$\theta_{JN}~({\rm rad})$ & $1.034$\\
$\alpha~({\rm rad})$ & $3.407$\\
$\delta~({\rm rad})$ & $0.401$\\
$\psi~({\rm rad})$ & $1.926$\\
Network optimal SNR & 20
\end{tabular}
\end{ruledtabular}
\end{table}

The mismatch between each candidate simulation and all recovery waveform
approximants is computed following the procedure described in
Sec.~\ref{sec:sm_match}.
The 40 simulations with the largest mismatch between the SXS waveform
and the \textsc{NRSur7dq4} approximant are selected as the final
injection set.
Their source properties are listed in Tab.~\ref{tab:sxs_pars}.
As described in the main text, only the 20 simulations with the largest mismatches (the
first 20 entries of Tab.~\ref{tab:sxs_pars}) are recovered under all three
hypotheses and reported in detail throughout this work. The remaining 20
simulations (the last 20 entries of Tab.~\ref{tab:sxs_pars}) are recovered under the
unlensed and phenomenological-lensing hypotheses only, as a check on the generality
of our results; since this check yields no new qualitative information, the
corresponding individual results are not reported, either in the main text or in
this Supplemental Material.

\begin{table*}[t]

\captionsetup{justification=raggedright,singlelinecheck=false}
	\caption{Binary parameters of the 40 selected SXS simulations. 
The table lists the SXS waveform identifiers, mass-ratio $q=m_2/m_1$, 
dimensionless spin magnitudes of the primary and secondary black holes 
$a_1$ and $a_2$, spin tilt angles relative to the orbital angular momentum 
$\theta_1$ and $\theta_2$ (in radians), relative spin azimuthal angle 
$\phi_{12}$ (in radians), and luminosity distance $d_L$ (in Mpc). The first 20 
entries have the largest \nrsur \, mismatches and are recovered under all three 
hypotheses in the main text; the last 20 entries are recovered under the unlensed 
and phenomenological-lensing hypotheses only (see main text).}

\label{tab:sxs_pars}
\begin{ruledtabular}
\begin{tabular}{cccccccc}
SXS ID       & q      & $a_1$  & $a_2$  & $\theta_1$ & $\theta_2$ & $\phi_{12}$ & $d_L \, (Mpc)$   \\
\hline 
4088 & 0.5000    & 0.8990  & 0.8994 & 1.3316   & 0.9273   & 6.2293   & 978.31  \\
4089 & 0.5001 & 0.8990  & 0.8992 & 1.4609   & 1.0905   & 5.7805   & 1749.94 \\
0752 & 0.5000    & 0.7993 & 0.7994 & 1.0404   & 1.0030    & 5.9133   & 1322.90  \\
0906 & 0.5000    & 0.7993 & 0.7994 & 1.5653   & 0.6121   & 6.0425   & 1002.36 \\
1282 & 0.5001 & 0.8789 & 0.8792 & 0.7976   & 1.3931   & 4.6375   & 2775.79 \\
4038 & 1.0000    & 0.9487 & 0.9487 & 0.6978   & 1.1836   & 5.6236   & 3192.42 \\
4095 & 0.5001 & 0.8989 & 0.8991 & 1.3801   & 1.6699   & 1.7807   & 1425.09 \\
0680 & 0.6000    & 0.7991 & 0.7994 & 0.7953   & 0.9764   & 0.0451   & 1594.64 \\
0750 & 0.5000    & 0.7994 & 0.7994 & 1.1267   & 2.0219   & 6.2080    & 1612.98 \\
0400 & 0.5715 & 0.7994 & 0.7992 & 1.2687   & 1.7871   & 5.9529   & 1079.28 \\
4285 & 0.5000    & 0.9487 & 0.8992 & 1.4205   & 1.8088   & 1.0905   & 1131.80  \\
0623 & 0.9000    & 0.8988 & 0.8988 & 1.4483   & 1.6417   & 6.2524   & 2501.16 \\
1200 & 0.5001 & 0.8491 & 0.8490  & 1.4341   & 1.8125   & 5.1067   & 1716.48 \\
4091 & 0.5000    & 0.8989 & 0.8991 & 1.3933   & 1.9089   & 5.0653   & 1976.22 \\
0716 & 0.5001 & 0.7994 & 0.7992 & 1.4447   & 1.7903   & 5.1177   & 1596.68 \\
0772 & 0.5000    & 0.7993 & 0.7994 & 1.3515   & 2.0644   & 4.7115   & 1241.38 \\
0696 & 0.6000    & 0.7991 & 0.7994 & 1.9367   & 0.8407   & 0.3639   & 1872.41 \\
0722 & 0.5001 & 0.7994 & 0.7994 & 1.4306   & 1.8586   & 5.0817   & 1316.00  \\
0917 & 0.5000    & 0.7990  & 0.7994 & 1.6005   & 1.5096   & 5.4330    & 2175.63 \\
4079 & 0.5001 & 0.8989 & 0.8992 & 0.7600     & 0.3800    & 2.4699   & 1733.92 \\
4093 & 0.5001 & 0.8989 & 0.8993 & 1.4278   & 0.7986   & 3.0855   & 1999.34 \\
0941 & 0.5000    & 0.7995 & 0.7993 & 0.6997   & 1.7793   & 5.8055   & 1403.26 \\
0776 & 0.5000    & 0.7991 & 0.7994 & 1.5503   & 0.1102   & 3.1247   & 1271.55 \\
0714 & 0.5000    & 0.7991 & 0.7995 & 1.5727   & 0.0820   & 3.1487   & 2809.07 \\
0763 & 0.5000    & 0.7991 & 0.7996 & 1.5733   & 0.0732   & 3.3257   & 2550.59 \\
0837 & 0.5000    & 0.7992 & 0.7996 & 1.5759   & 0.0722   & 3.2749   & 2602.11 \\
3423 & 0.5806 & 0.7914 & 0.7599 & 0.9871   & 1.3730    & 5.7519   & 2179.55 \\
0886 & 0.5000    & 0.7992 & 0.7994 & 1.4990    & 0.8204   & 4.4972   & 1526.54 \\
0888 & 0.5000    & 0.7990  & 0.7994 & 1.598    & 0.5189   & 6.0182   & 2249.50  \\
0896 & 0.5000    & 0.7991 & 0.7996 & 1.5508   & 0.8784   & 2.7247   & 1872.41 \\
0654 & 0.7500   & 0.7990  & 0.7995 & 1.0801   & 0.6359   & 5.7733   & 2134.47 \\
0932 & 0.5000    & 0.7995 & 0.7994 & 0.5673   & 1.1986   & 0.2149    & 1673.37 \\
0705 & 0.5000    & 0.7995 & 0.7994 & 0.5597   & 1.2047   & 0.2556   & 1682.90  \\
0980 & 0.5000    & 0.7994 & 0.7994 & 0.5534   & 1.2059   & 0.2767   & 1761.09 \\
0463 & 0.5129 & 0.7909 & 0.7188 & 1.1183   & 2.1103   & 0.0929   & 2346.15 \\
0814 & 0.5000    & 0.7993 & 0.7994 & 1.3458   & 2.0846   & 4.3860    & 1354.56 \\
0968 & 0.5000    & 0.7992 & 0.7996 & 1.6474   & 1.2556   & 0.0374   & 1872.41 \\
0991 & 0.5000    & 0.7990  & 0.7994 & 1.5901   & 1.5264   & 5.3903   & 2216.10  \\
1215 & 0.5000    & 0.8490  & 0.8493 & 1.4226   & 1.8799   & 5.0870    & 1303.00  \\
0977 & 0.5001 & 0.7993 & 0.7993 & 1.3585   & 2.0743   & 4.4954   & 2063.13 \\ 
\end{tabular}
\end{ruledtabular}
\end{table*}

\section{Waveform mismatch calculation}
\label{sec:sm_match}
The waveform mismatch used to rank candidate injections is computed using the
generalized-match procedure of Refs.~\cite{Harry_2016,Hamilton_2021}.
The noise-weighted inner product is defined as~\cite{Veitch:2014wba}

\begin{equation}
\langle a|b\rangle
= 4 \, {\rm Re}
\int_{f_{\rm low}}^{f_{\rm high}}
\frac{\tilde a(f)\tilde b^*(f)}
{S_n(f)},df,
\label{eq:inner_product_sm}
\end{equation}

where $S_n(f)$ is the noise power spectral density (PSD). Given an SXS signal $h_s$ and a template $h_t$ generated with the same intrinsic parameters, 
the match is defined as 
\begin{equation}
    m= \max_{t_c,\phi_c,\psi,\alpha_0} \frac{\langle h_s|h_t\rangle} 
    {\sqrt{\langle h_s|h_s\rangle \langle h_t|h_t\rangle}}, 
    \label{eq:match_def_sm} 
\end{equation} 

and the mismatch is $\mathcal{M}=1-m$. 
The maximization is performed over the coalescence time $t_c$, the coalescence phase $\phi_c$, and the
template polarization angle $\psi$, following the analytic and FFT-based procedure of Ref.~\cite{Harry_2016}, and additionally over the 
initial in-plane spin orientation of the template, $\alpha_0$, following Ref.~\cite{Hamilton_2021}; full details of the optimization are given therein.
We use a low-frequency cutoff of 20~Hz and the L1 noise PSD
adopted in the waveform-systematics analysis of GW231123~\cite{LIGOScientific:2025rsn}.

\section{Prior distributions}
\label{sec:sm_priors}

This section summarizes the prior distributions adopted for the Bayesian
parameter estimation analyses. For the unlensed analyses, the chirp mass is sampled uniformly in the range
$100\, \msun \le \mathcal{M}_c \le 180\, \msun $, and the mass ratio
is drawn uniformly in component masses corresponding to
$q \in [1/6, 1]$, with constraints on primary and secondary masses $m_i \in [1,1000]$. 
Spin magnitudes are sampled uniformly in $[0,0.99]$ , while spin
orientations are taken to be isotropic, with tilt angles following
sine distributions. The sky location is assumed isotropic, the
polarization angle is uniform in $[0,\pi]$, and the coalescence
phase is uniform in $[0,2\pi]$.
The luminosity distance is sampled uniformly in comoving volume
between $10\,\mathrm{Mpc}$ and $10000\,\mathrm{Mpc}$ assuming $\Lambda$CDM cosmology. The coalescence
time is uniform in a narrow window of $200\,\mathrm{ms}$ around the
injection time, taken to coincide with the GW231123 trigger time~\cite{LIGOScientific:2025rsn}.

For the lensing analyses, the prior distributions for the isolated point-mass (PM) lens models 
are chosen to match those used in Ref.~\cite{LIGOScientific:2025cwb}, while the priors for the phenomenological lensing analyses are similar to those in ~\cite{PhenomPaper}.
For the PM model, the two additional degrees of freedom are the redshifted lens mass $M_{l}^{z}$, sampled uniformly in logarithmic space in $[1, 10^4]M_\odot$, and the source position 
$y$ drawn from a power-law distribution with index $1$ over $[0.01, 3]$. 
For the phenomenological lens model, we allow for 6 overlapping images linked via
a time delay, an overall magnification and a possible phase shift. 
The relative magnifications from one image to another are sampled from 
a log-uniform distribution over $[0.01, 100]$, and the inter-image time delays 
are sampled uniformly in the range $[0.001, 1]\,\mathrm{s}$. 
Each image $i$ is also assigned a Morse phase index $n_i$, drawn from a 
discrete uniform distribution over allowed values of $\{0,1,2\}$. 

\section{Additional posterior distributions}
\label{sec:sm_posteriors}

This section presents posterior distributions for all 20 injections analyzed
in this work, excluding the four case studies discussed in the main text.
For many injections, the lensed and unlensed recoveries produce nearly
identical posterior distributions for the intrinsic binary parameters,
indicating that the additional lensing degrees of freedom do not affect the inference.
Noticeable deviations occur only for a subset of injections,
including those discussed individually in the main text, where waveform
systematics produce a stronger preference for lensing models. Figs.~\ref{fig:phenom_posteriors_1}--
\ref{fig:pm_posteriors_2}
show normalized posterior probability density histograms for the
intrinsic binary parameters recovered under the unlensed,
phenomenological, and PM lensing hypotheses.

\section{Lens parameter posteriors}
\label{sec:sm_lens}

Figs.~\ref{fig:lens_posteriors_1} and \ref{fig:lens_posteriors_2}
summarize the posterior distributions of the lens-model parameters for
all injections, excluding the four case studies discussed in the main text.

\begin{figure*}
\captionsetup{justification=raggedright,singlelinecheck=false}
\includegraphics[width=\textwidth]{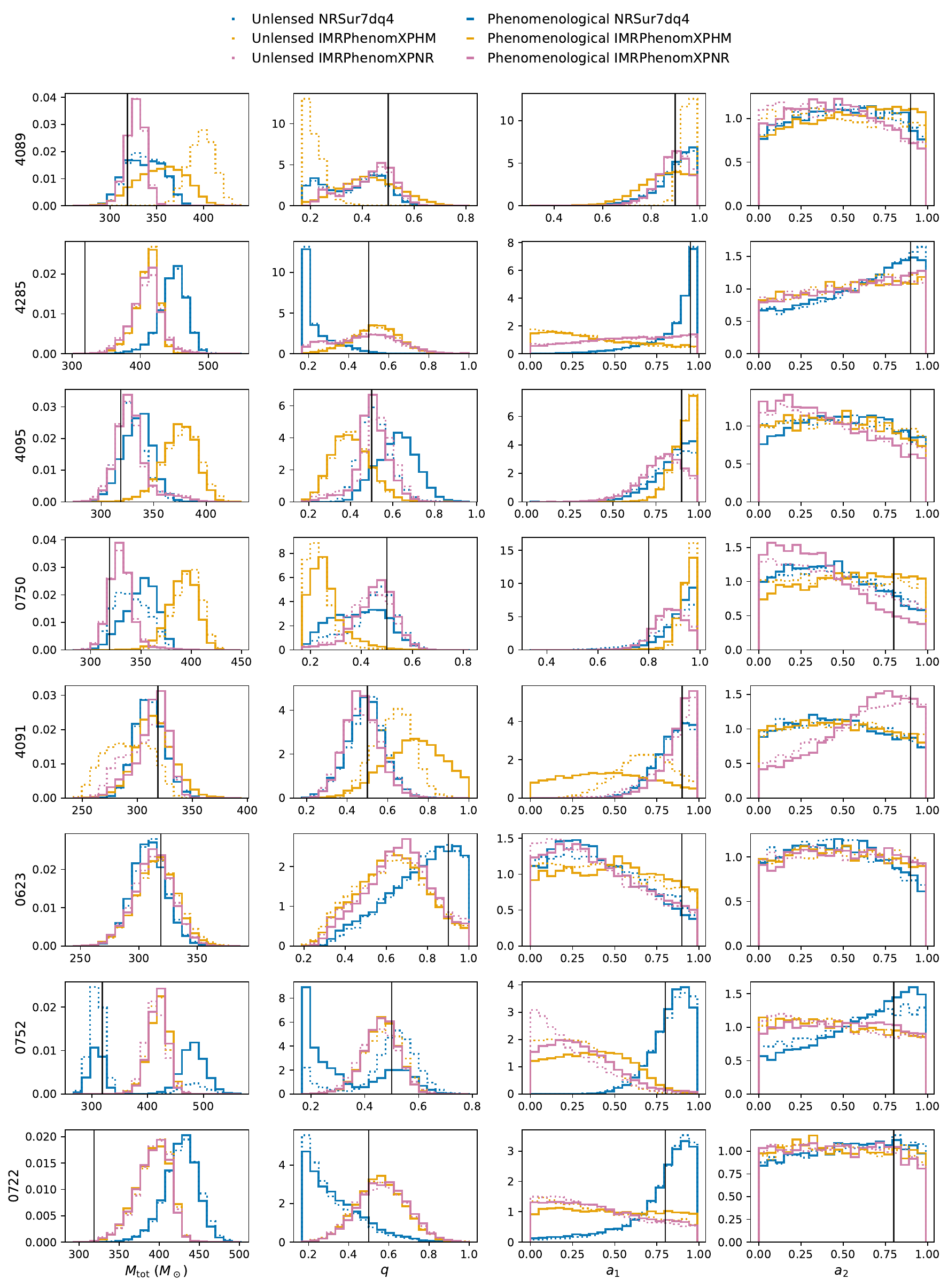}
\caption{
Posterior probability density functions (1D marginal distributions) for BBH 
parameters across different SXS injections, comparing unlensed (dotted) and phenomenologically 
lensed (solid) recoveries. Black vertical lines represent the injected values.
}
\label{fig:phenom_posteriors_1}
\end{figure*}

\begin{figure*}
    \captionsetup{justification=raggedright,singlelinecheck=false}
    \includegraphics[width=\textwidth]{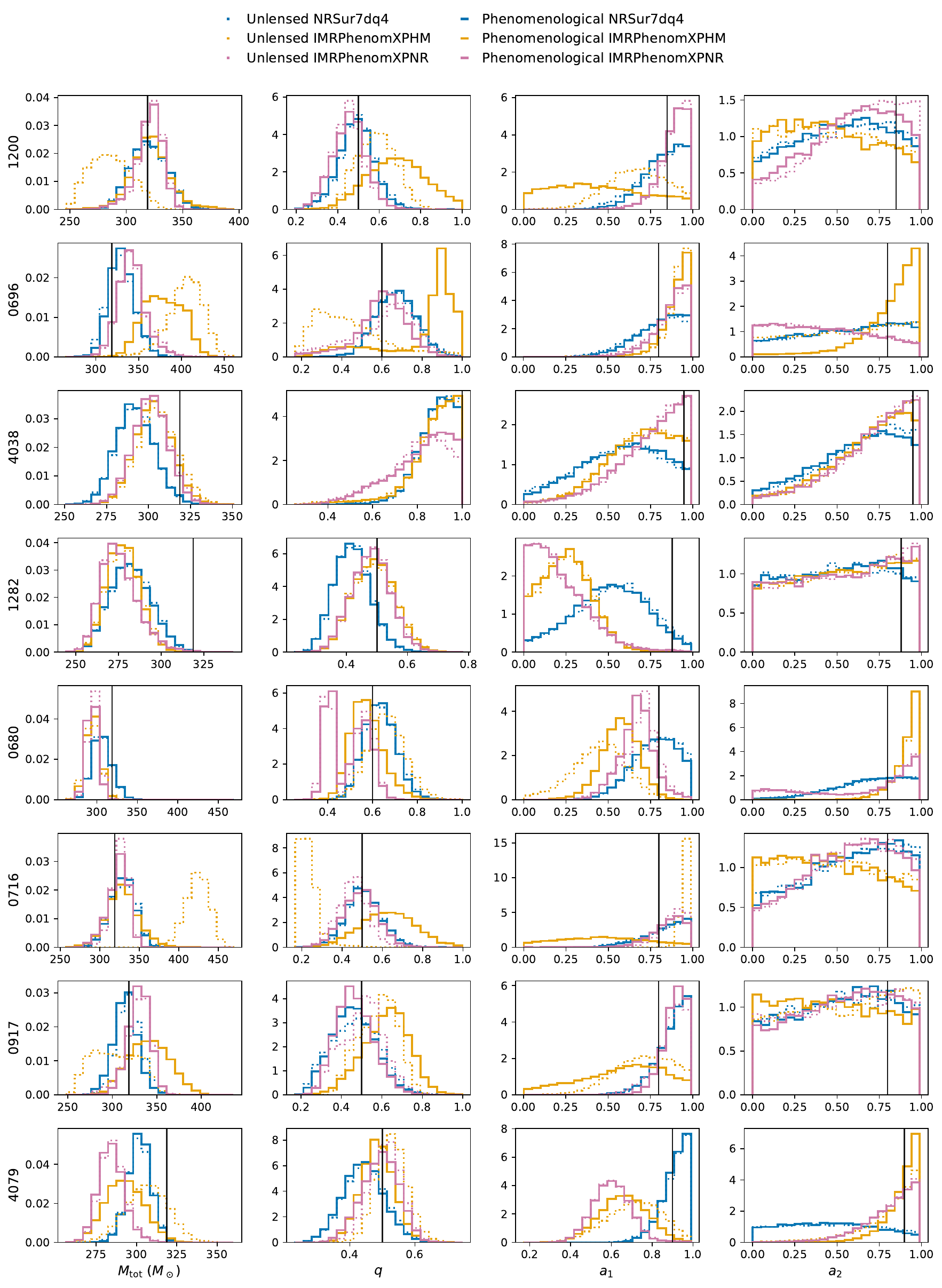}
    \caption{Continued: same as Fig.~\ref{fig:phenom_posteriors_1} for remaining injections.}
    \label{fig:phenom_posteriors_2}
\end{figure*}

\begin{figure*}
\captionsetup{justification=raggedright,singlelinecheck=false}
\includegraphics[width=\textwidth]{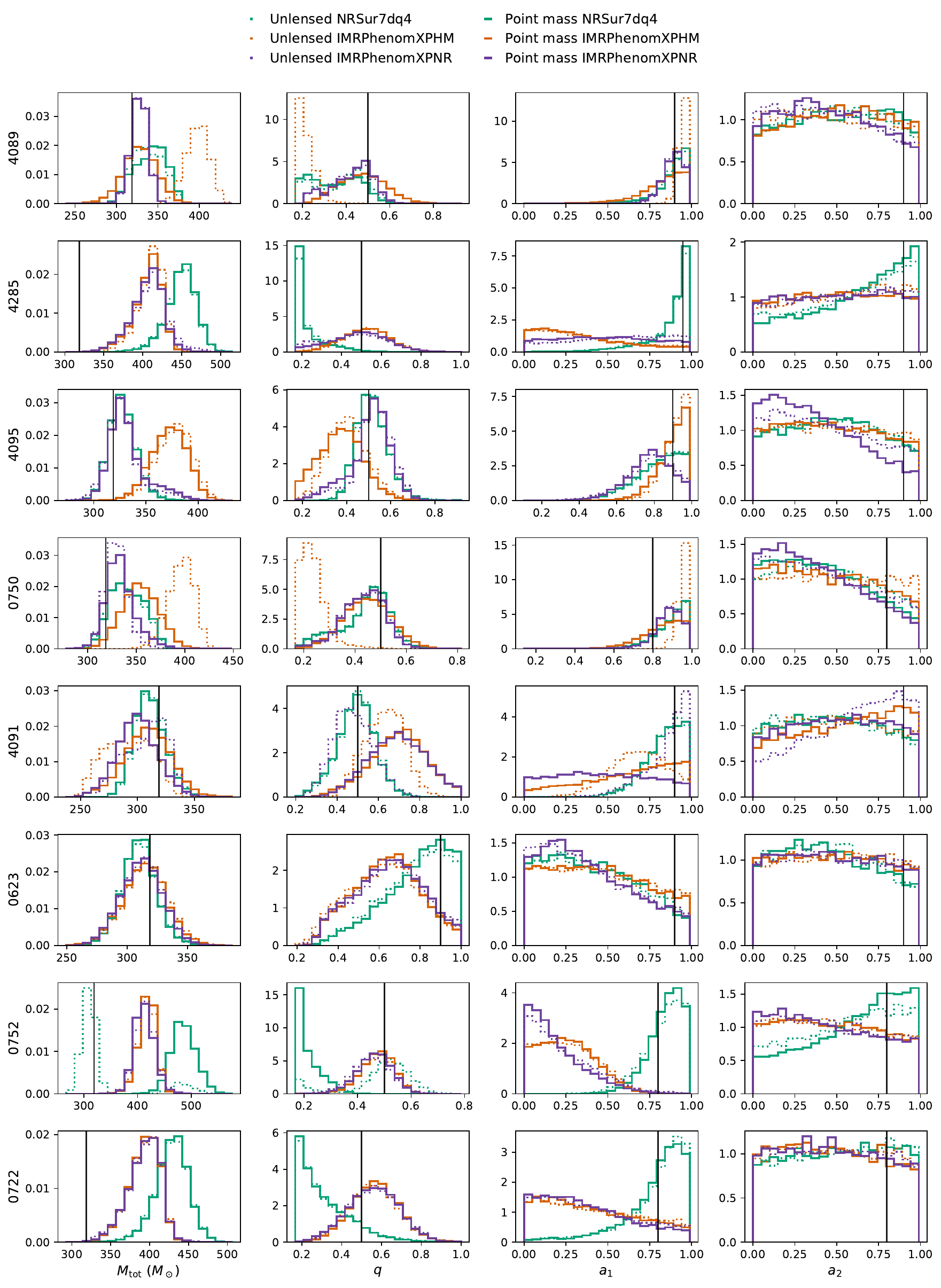}
\caption{
Posterior probability density functions for BBH
parameters across different SXS injections, comparing unlensed (dotted) and point mass
lensed (solid) recoveries. Black vertical lines represent the injected values.
}
\label{fig:pm_posteriors_1}
\end{figure*}

\begin{figure*}
\captionsetup{justification=raggedright,singlelinecheck=false}
\includegraphics[width=\textwidth]{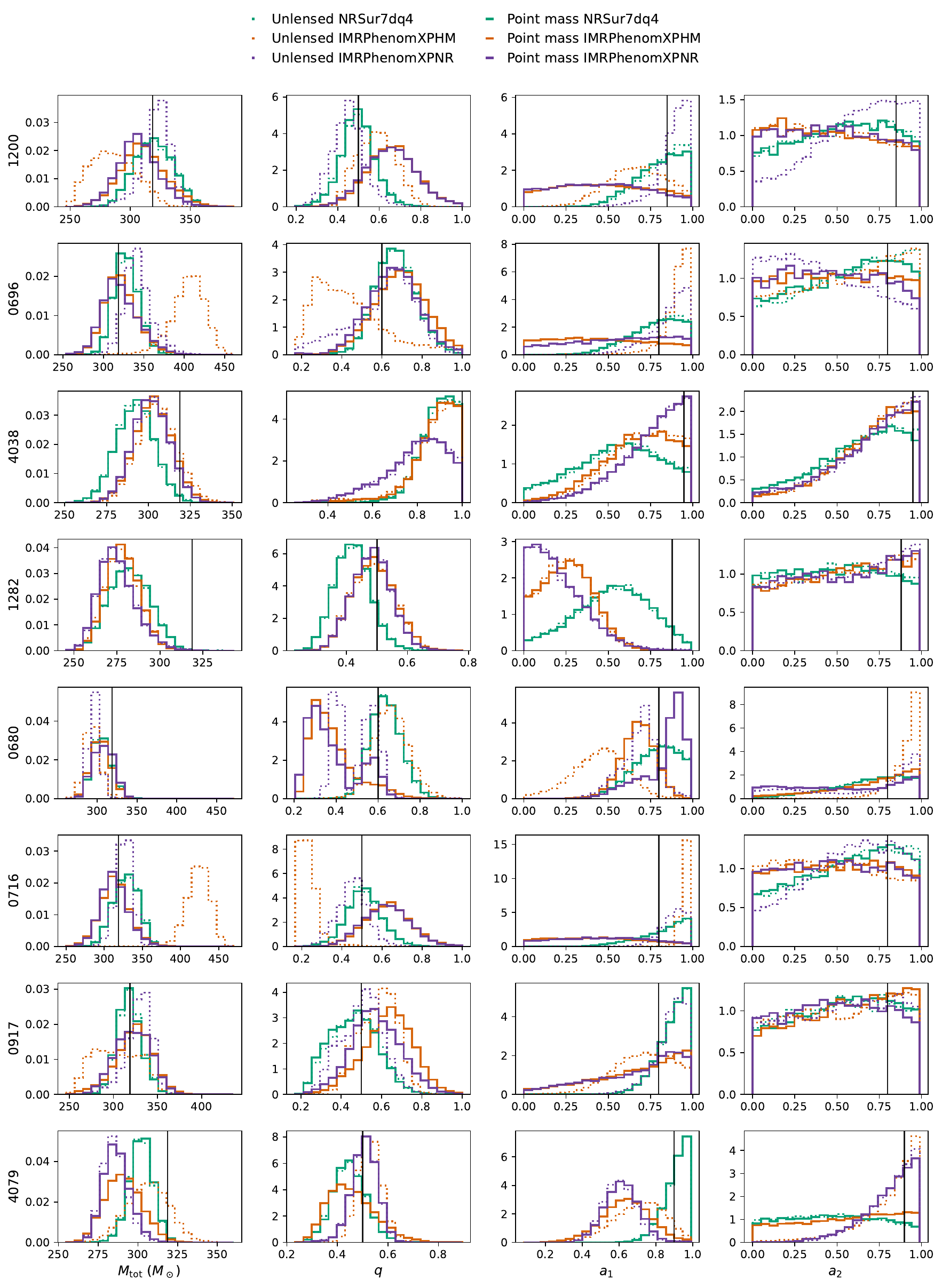}
\caption{Continued: same as Fig.~\ref{fig:pm_posteriors_1} for remaining injections.
}
\label{fig:pm_posteriors_2}
\end{figure*}

\begin{figure*}[t]
\captionsetup{justification=raggedright,singlelinecheck=false}
\includegraphics[width=\textwidth]{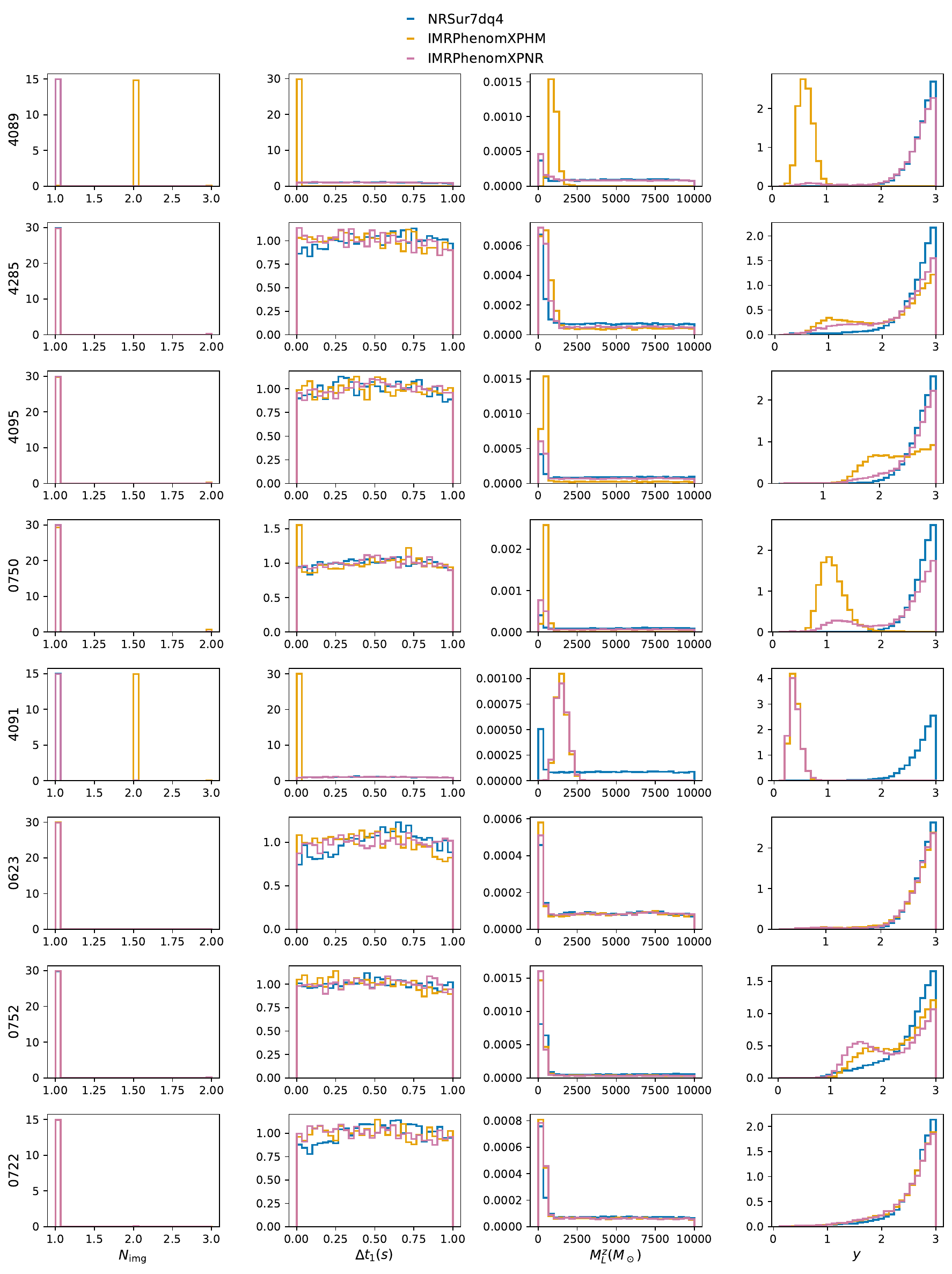}
\caption{
Posterior probability density functions of lens parameters recovered from the SXS injections.
Columns show (from left to right) the number of lensed images $N_{\mathrm{img}}$ and the time delay between the first two images $\Delta t_1$, inferred from the phenomenological model, the redshifted lens mass $M_L^{z}$, and the source position $y$ inferred from the point mass lens model. 
}
\label{fig:lens_posteriors_1}
\end{figure*}

\begin{figure*}[t]
\captionsetup{justification=raggedright,singlelinecheck=false}
\includegraphics[width=\textwidth]{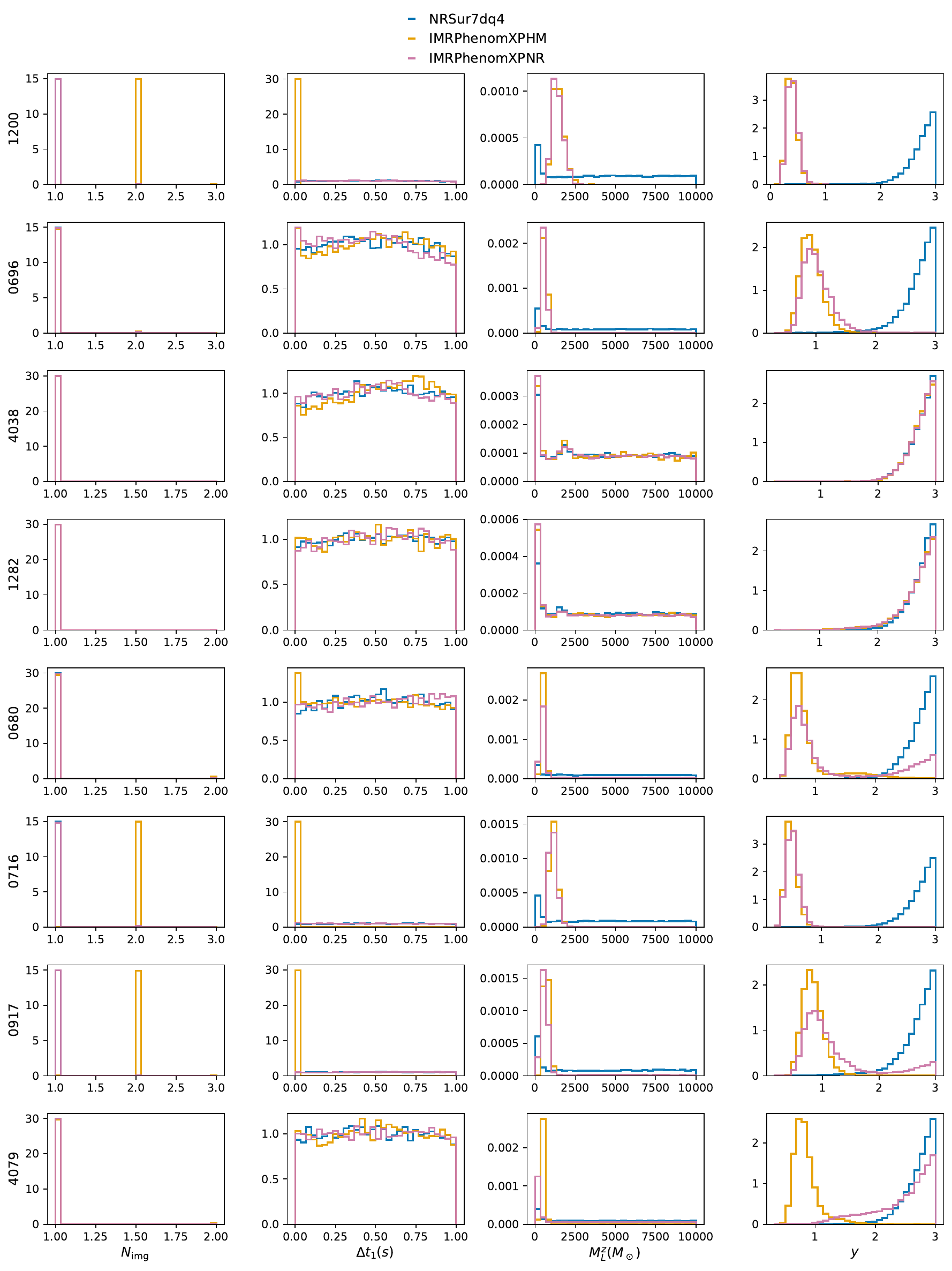}
\caption{Continued: same as Fig.~\ref{fig:lens_posteriors_1} for remaining injections.
}
\label{fig:lens_posteriors_2}
\end{figure*}

\bibliographystyle{apsrev4-2}
\bibliography{references}